\documentclass[aps,pra, twocolumn, nofootinbib, showpacs]{revtex4-1} 
\usepackage[unicode=true,pdfusetitle,
bookmarks=true,bookmarksnumbered=false,bookmarksopen=false,
breaklinks=false,pdfborder={0 0 0},backref=false,colorlinks=false] {hyperref}
\hypersetup{ colorlinks,linkcolor=myurlcolor,citecolor=red,urlcolor=myurlcolor}
\usepackage[T1]{fontenc}

\usepackage{braket,colortbl,amsthm,amsmath,cleveref,amssymb,txfonts}\definecolor{myurlcolor}{rgb}{0,0,0.7}
\usepackage{graphics,graphicx}
\usepackage{color}
\usepackage{amssymb}
\usepackage{amsthm}
\usepackage{amsfonts}
\usepackage{float}
\usepackage{graphicx}
\usepackage[format = plain,labelfont = bf,up, textfont = normal , up, justification
=raggedright, singlelinecheck =false]{caption}
\usepackage{subcaption}

\newcommand\norm[1]{\left\lVert#1\right\rVert}
\newcommand \modu[1]{ \lvert#1\rvert}
\usepackage{tabularx}
\usepackage{amsmath}
\usepackage{braket}

\usepackage{graphicx}
\usepackage[utf8x]{inputenc}
\usepackage{color,soul}
\usepackage{amsmath}
\usepackage{braket}
\usepackage{latexsym}
\usepackage{amssymb}
\usepackage{amsthm}
\usepackage{xcolor}
\usepackage{bm}
\usepackage{graphics,epstopdf}
\usepackage{color}\usepackage{amsmath}
\usepackage{enumitem}
\usepackage{fmtcount}
\usepackage{booktabs}
\usepackage{hyperref}
\usepackage{epsfig}

\theoremstyle{plain}

\begin{document}

		\title{Convolution algebra of superoperators and nonseparability witnesses for quantum operations}
	\author{Sohail}
	\email{sohail@hri.res.in}
	\author{Ujjwal Sen}
	\email{ujjwal@hri.res.in}
	\affiliation{Harish-Chandra Research Institute,  A CI of Homi Bhabha National
Institute, Chhatnag Road, Jhunsi, Prayagraj - 211019, India}
\begin{abstract}
     We define a product between quantum superoperators which is preserved under the Choi-Jamio{\l}kowski-Kraus-Sudarshan channel-state isomorphism. We then identify the product as the convolution on the space of superoperators, with respect to which the channel-state duality is also an algebra isomorphism. We find that any witness operator for detecting nonseparability of quantum operations on separated parties can be written entirely within the space of superoperators with the help of the convolution product.
\end{abstract}
	\maketitle
\section{Introduction} 
For isolated physical systems, the formalism of quantum mechanics 
associates a separable Hilbert space 
to the system and the states are represented by  positive trace-class operators on that Hilbert space  with unit trace. Observables are represented by hermitian operators and time evolution is described by unitary operators on the same space. In a realistic scenario, we can never have a completely isolated physical system, as there will always be interactions with the environment. 
One way to treat such a system is to consider the system and its environment jointly as an isolated system. So the time evolution of the joint system is a unitary time evolution. We then trace out the environment part in order to get the evolution of the system. This  leads to the formalism of quantum operations.
Mathematically, quantum operations are linear maps on the algebra of bounded operators on the Hilbert space associated with the physical system. As quantum operations map a state to another state (i.e., it maps density matrices to density matrices) of the physical system, it is required to be positivity and trace preserving.
However, since one can apply a physical operation on a part of a physical system, a valid physical operation also needs to be completely positive, i.e., it must preserve the positivity of states of all extensions of the system to bigger ones, for which tracing out the extra parts, one regains the original state of the smaller system. 

The positivity and complete positivity of linear maps between matrix algebras  are therefore crucial concepts in quantum theory as just in operator algebra.
Let \(\mathcal{B}(\cdot)\) denote the space of bounded operators on the Hilbert space in the argument.
A map $\phi : \mathcal{B}(H_1) \rightarrow \mathcal{B}(H_2)$ is positive if it takes any positive element of $\mathcal{B}(H_1)$ to a positive element of $\mathcal{B}(H_2)$. Now $\phi$ induces a map $\phi^{(k)}:M_k (\mathcal{B}(H_1)) \rightarrow M_k (\mathcal{B}(H_2))$, where $M_k (\mathcal{B}(\cdot))$ is a $k \times k$ matrix whose entries are elements of $\mathcal{B}(\cdot)$. $\phi$ is called \(k\)-positive if $\phi^{(k)}$ maps positive elements of $M_k (\mathcal{B}(H_1))$ to positive elements of $M_k (\mathcal{B}(H_2))$. If $\phi$ is \(k\)-positive for all $k \in \mathbb{N}$, then we call it a completely positive map. Note that while complete positivity implies positivity,  the converse is not true in general. An example of a map which is positive but not complete positive is the transposition of matrices. This property of transposition is used to detect whether certain shared quantum states are entangled \cite{PhysRevLett.77.1413, 1996PhLA..223....1H}. 

The positive operators on a separable Hilbert space represent  quantum states of the corresponding physical system and the completely positive maps on the state space represent  physical operations on the quantum system. These two important concepts are connected to each other through The Choi-Jamio{\l}kowski-Kraus-Sudarshan  (CJKS) isomorphism~\cite{book,article,CHOI1975285,kraus1983states,Sudarshan1985}. It helps us to recognize quantum operations on the state space of a physical system as a quantum states on a bigger Hilbert space. It is a useful tool for studying the behavior of completely positive maps. The  CJKS isomorphism is the vector space isomorphism between the linear space $\mathcal{L}(\mathcal{B}(H_1),\mathcal{B}(H_2))$  of linear maps from $\mathcal{B}(H_1)$ to $\mathcal{B}(H_2)$ and the linear space $\mathcal{B}(H_1) \otimes \mathcal{B}(H_2)$. It does not preserve the algebra. Let $f$ represent the CJKS isomorphism and $\phi_1$ and $\phi_2$ be two maps in $\mathcal{L}(\mathcal{B}(H_1),\mathcal{B}(H_2))$. Then $f(\phi_1 \phi_2) \neq f(\phi_1) f(\phi_2)$, where the product used on both sides of the relation is the usual composition of linear maps.

In this work, we introduce a product in $\mathcal{L}(\mathcal{B}(H_1),\mathcal{B}(H_2))$ which is preserved under the CJKS isomorphism, so that the latter turns into an algebra isomorphism. The product has clear similarities with a convolution. We discuss these similarities and other properties of the convolution product. We then use it to build the apparatus of witnessing the nonseparability of quantum operations on shared systems entirely staying within the space of superoperators.  
We exemplify the apparatus by using it on the controlled-NOT, swap, and the controlled-Z gates on the space of two qubits.
 
\section{The CJKS Isomorphism}

Let \(H_1\) and \(H_2\) be two finite-dimensional Hilbert spaces, 
and let $\phi : \mathcal{B}(H_1)\rightarrow \mathcal{B}(H_2)$ be a linear map, where 
\(H_1 \equiv \mathbb{C}^n\).
Let $ \{e_{ij}\}$, $i,j=1,2,\ldots,n$ be a complete set of matrix units for $\mathcal{B}(H_1)$. Then the CJKS  matrix \cite{book,article,CHOI1975285,kraus1983states,Sudarshan1985} for $\phi$ is defined as the operator
        $\rho_\phi =  \sum_{{i,j}=1}^{n} $ $e_{ij} \otimes \phi(e_{ij})$ $\in $ $\mathcal{B}(H_1)\otimes \mathcal{B}(H_2) $.
 The map $\phi \rightarrow \rho_\phi $ is linear and bijective, and is called the CJKS isomorphism. The  isomorphism leads to the concept of a ``channel-state duality". Here, a channel or quantum channel is a completely positive trace-preserving map, which acts on the space of bounded operators on a Hilbert space. To understand the ``channel-state duality" we need to have a look at the CJKS theorem on completely positive maps.\\
 

\noindent \textbf{CJKS theorem on completely positive maps~\cite{book,article,CHOI1975285,kraus1983states,Sudarshan1985}.} \emph{The CJKS matrix $\rho_\phi = $ $\sum_{{i,j}=1}^{n} $ $e_{ij} \otimes \phi(e_{ij})$ $\in $ $\mathcal{B}(H_1)\otimes \mathcal{B}(H_2) $ is positive if and only if the map $\phi : \mathcal{B}(H_1)\rightarrow \mathcal{B}(H_2)$ is completely positive.}\\

The CJKS isomorphism, with the help of the CJKS theorem on completely positive maps, allows us to view completely  positive trace-preserving linear maps acting on quantum states as a quantum state in a higher-dimensional Hilbert space. If we consider quantum states which are density matrices on an $n$-dimensional Hilbert space, then the completely positive trace-preserving map acting on them can be identified with a density matrix on an  $n^2$-dimensional Hilbert space.



\section{CJKS isomorphism as  algebra isomorphism}

Let $\phi  \in \mathcal{L}(\mathcal{B}(H_1),\mathcal{B}(H_2))$, 
the set of all linear maps from $\mathcal{B}(H_1)$ to $\mathcal{B}(H_2)$, for two Hilbert spaces $H_1$ and $H_2$. The map  $f:\mathcal{L}(\mathcal{B}(H_1),\mathcal{B}(H_2)) \rightarrow \mathcal{B}(H_1)\otimes \mathcal{B}(H_2) $, defined by
$f(\phi)=\rho_\phi=\sum_{{i,j}=1}^{n} $ $e_{ij} \otimes \phi(e_{ij})$, is the CJKS isomorphism. 
         
 The set, $\mathcal{B}(H)$, for a Hilbert space, \(H\), is an algebra where  the ``multiplication'' of $T_1$, $T_2$ $\in$ $ \mathcal{B}(H)$ is taken as the usual composition of two maps, i.e., the one defined via \((T_1 T_2)(x):= T_1(T_2 x)\) for all $x \in H$. We will consider the algebra, $\mathcal{B}(H_1)\otimes \mathcal{B}(H_2)$, 
 where 
 the said multiplication is defined via 
$(T_1 \otimes T_2)(T_1{'} \otimes T_2{'})=(T_1 T_1{'})\otimes (T_2 T_2{'}) $, where 
 $T_1$, $T_1{'}$ $\in$ $ \mathcal{B}(H_1)$ and $T_2$, $T_2{'}$ $\in$ $ \mathcal{B}(H_2)$.


We now 
define a  multiplication in $\mathcal{L}(\mathcal{B}(H_1),\mathcal{B}(H_2))$, that is different  from and ``additional'' to the usual composition in the same space, 
in the following way. We denote the additional   ``multiplication" by \(*\), and define 
\begin{equation}
\phi_1 * \phi_2(e_{ij}):= \sum_{k} \phi_1(e_{ik}) \phi_2(e_{kj}), 
\label{gubleT}
\end{equation}
with the action on other arguments being obtained by linearity. 
It is important to note that the word ``additional'' does \emph{not} imply that an algebra has two multiplications defined on it.  The set $\mathcal{L}(\mathcal{B}(H_1),\mathcal{B}(H_2))$ forms two separate algebras with respect to the two compositions. 
We can easily verify that $\mathcal{L}(\mathcal{B}(H_1),\mathcal{B}(H_2))$ is an associative algebra with the multiplication, \(*\),
and with respect to this multiplication, the CJKS isomorphism is not only a vector space isomorphism, it is also an \emph{algebra isomorphism}, i.e., \begin{equation}
f(\phi_1 * \phi_2)=f(\phi_1)f(\phi_2),
\end{equation}
where the multiplication on the right hand side is the multiplication in $\mathcal{B}(H_1)\otimes \mathcal{B}(H_2)$. Furthermore, it is an algebra with identity, with the identity element being
\begin{center}
$\phi_e(e_{il})=\delta_{il} \mathbb{I}$,
\end{center}
extended to other arguments by linearity.








Let us \textbf{digress} a little, and discuss about the convolution of functions, and its similarity with the additional multiplication defined above.
\\\\
\textbf{Convolution algebra.} An algebra $\mathbb{A}$ over a field $F$ is a vector space $\mathbb{A}$ over $F$, where a rule for multiplying two vectors is defined in such a way that the multiplication is associative, i.e., $x(yz)=(xy)z$, distributive over the addition of vectors from both the sides, i.e., $x(y+z)=xy+xz$ and $(y+z)x=yx+zx$, and scalar multiplication follows $\alpha(xy)=(\alpha x)y=x(\alpha y)$, for all $\alpha \in F$ and for all $x,y,z \in \mathbb{A}$. An algebra may or may not contain a multiplicative identity element, but if it does, then it is called a unital algebra. The set of real functions defined on the real line is a vector space over the field of real numbers under usual addition and scalar multiplication. It is easy to see that this vector space is an algebra over real numbers under the composition of functions. But composition is not the only multiplication which can be defined on the space of real functions to make it an algebra. The convolution of two real (Lebesgue) integrable functions is another kind of multiplication that can be defined on the space of integrable functions. If $f_1:\mathbb{R} \rightarrow \mathbb{R}$ and $f_2: \mathbb{R} \rightarrow \mathbb{R}$ are two real functions, then their convolution 
is defined as $f_1 * f_2(x)= \int _{\mathbb{R}} f_1(y)f_2(x-y)dy$. Convolution makes the linear space of integrable functions an 
algebra. The definition can of course be 
 generalized to multi-variable functions.
 The convolution integral does not always exist.
 
 Convolutions can also be defined for functions on groups, and let us discuss about finite groups only. 
%
%
Let $G$ be a (finite) group and $F$ be a field,
and let $L(G,F)$ be the vector space of  maps from $G$ to $F$. The convolution of $f_1, f_2 \in L(G,F)$ is defined as 
\begin{equation}
    f_1 * f_2 (g) = \sum_{xy=g} f_1(x)f_2(y) \label{conv}
\end{equation}
 $\forall g \in G$, where the sum is over the pairs $(x,y)$ from \(G\) such that $xy=g$.
Equivalently, the convolution can be expressed as 
\begin{equation}
    f_1 * f_2 (g) = \sum_{x} f_1(x)f_2(x^{-1} g).
\end{equation}
The convolution operation $*$ makes $L(G,F)$ an algebra, known as the convolution algebra. \\
\\
\textbf{Group Algebra.} Let $G$ be a group of cardinality $d$. The group algebra of $G$ over the field of complex numbers is an 
algebra whose elements are  linear combinations of the elements of $G$, i.e., $G$ is a basis for this algebra. An arbitrary element $a$ of the group algebra looks like $a= \sum_{i=1} ^d \alpha_i g_i$, where $\alpha_i$'s are complex numbers. The product in the group algebra is induced by the group multiplication. If $a= \sum_{i=1} ^d \alpha_i g_i$ and $b= \sum_{i=1} ^d \beta_j g_j$ are two elements of the group algebra then the multiplication of these elements is given by $a\cdot b=\sum_{i=1}^d \sum_{j=1}^d \alpha_{i} \beta_{j} g_{i} g_{j} $. The identity element of the algebra is the identity element of the group.

It can be shown that the convolution algebra $L(G,F)$ is isomorphic to the group algebra of $G$. 
(For details about convolution algebra, group algebra, and related topics, the Reader may have a look at \cite{1967RvMP...39..259L}.)
Directions to the proof are given in the following discussion, which also helps to set up some notations and terminology.

Let the cardinality of $G$ be $d$, and let us only focus on complex fields. A functional $f:G \rightarrow \mathbb{C}$ can be completely  specified by its action on each element of $G$, i.e., by specifying the complex numbers $f(g_i)$ for all $i$ from 1 to $d$. So $f$ is completely specified by the $d$-tuple $(f(g_1), f(g_2),....,f(g_d) )$. The set $ L(G,\mathbb{C})$ of functionals is a vector space under the usual operations, $(f_1+f_2)(g):=f_1(g)+f_2(g)$ and $(\alpha f)(g):=\alpha f(g)$. It is a vector space of dimension $d$ and we can take the set $\{ e_i \vert e_i (g_j) = \delta_{ij} \}$ as the basis for it.

Let $\mathcal{A}$ denote the group algebra of $G$ over the complex field $\mathbb{C}$.  A generic element $a \in \mathcal{A}$ has the form $a= \sum_{i=1} ^d \alpha_i g_i$.
  So, under the identification $a \leftrightarrow ( \alpha_1, \alpha_2,....,\alpha_d ) $, the two spaces $\mathcal{A}$ and $L(G,\mathbb{C})$ are isomorphic as vector spaces. Now there is a natural multiplication defined on $\mathcal{A}$, induced by the group multiplication, namely $a_1\cdot a_2 := \sum_{ij=1}^d f_1(g_i)f_2(g_j) g_i g_j$. Using the fact that in the group multiplication table, along a row or column each element of the group appears once and only once, the above sum can be rewritten as $a_1\cdot a_2 := \sum_{k=1}^d \sum_{i=1}^d f_1(g_i)f_2(g_i ^{-1} g_k) g_k=\sum_{k=1}^d  (f_1*f_2)( g_k) g_k$, where $(f_1*f_2)(g_k):= \sum_{i=1}^d f_1(g_i)f_2(g_i ^{-1} g_k)$ is  the convolution of $f_1$ and $f_2$. 
  This completes the proof that the convolution algebra $L(G,\mathbb{C})$ is isomorphic to the group algebra $\mathcal{A}$.

Consider now a group $G$ and an algebra $\mathcal{\Tilde{A}}$, and let us denote by $\mathcal{A}$  the group algebras of $G$. A  map $\phi : G \rightarrow \Tilde{\mathcal{A}}$  can be completely  specified by the tuple $(\phi(g_1), \phi(g_2),....,\phi(g_d) )$. Let  $L(G,\Tilde{\mathcal{A}})$ denote the set of all such maps. This is a vector space over $\mathbb{C}$ under the usual addition and scalar multiplication. Now the map $\phi$ can be extended to the map $\phi: \mathcal{A} \rightarrow \Tilde{\mathcal{A}}$ 
by linearity. The spaces $L(G,\Tilde{\mathcal{A}})=L(\mathcal{A},\Tilde{\mathcal{A}})$ and $\mathcal{A} \otimes \Tilde{\mathcal{A}}$ are isomorphic as vector spaces under the identification $(\phi(g_1), \phi(g_2),....,\phi(g_d) ) \leftrightarrow \sum_i g_i \otimes \phi(g_i)$ or equivalently by $\phi \leftrightarrow \sum_i g_i \otimes \phi(g_i)$. 

The spaces, $\mathcal{A} \otimes \Tilde{\mathcal{A}}$, with the multiplication rule induced by  group multiplication in $G$ and multiplication rule in $\Tilde{\mathcal{A}}$, and  $L(\mathcal{A},\Tilde{\mathcal{A}})$, with the convolution $\phi * \phi'(g_k)=\sum_i \phi(g_i) \phi'(g_i ^{-1} g_k)$, are algebras. Once again, with the help of  the fact that in the group multiplication table, along a row or column, each element of the group appears once and only once, it can  be proved the algebras  are isomorphic.

The convolution as defined in Eq.~(\ref{conv}) is applicable not only on functions over a group but also on functions over any algebraic structure with a binary operation. The following paragraph reflects this idea.


 Let us now \textbf{revert} back to 
 the space, $\mathcal{L}(\mathcal{B}(H_1),\mathcal{B}(H_2))$, of linear maps $\phi :\mathcal{B}(H_1) \rightarrow \mathcal{B}(H_2)$, and the tensor product, $\mathcal{B}(H_1) \otimes \mathcal{B}(H_2)$. A basis for $\mathcal{B}(H_1)$ is given by the matrix units, $\{ e_{ij} \}_{i,j=1}^n$.  
 The CJKS isomorphism, $\phi \leftrightarrow \sum_{{i,j}=1}^{n}  e_{ij} \otimes \phi(e_{ij}) $, 
 shows that the spaces $\mathcal{L}(\mathcal{B}(H_1),\mathcal{B}(H_2))$ and 
 $\mathcal{B}(H_1) \otimes \mathcal{B}(H_2)$ are isomorphic.
 It is obvious that the set $\{ e_{ij} \}_{i,j=1}^n$ does not form a group under matrix multiplication. So, $\mathcal{B}(H_1)$ is not a group algebra. But the set $\{ e_{ij} \}_{i,j=1}^n$ has the ``nice property'' that if we arrange them in a multiplication table similar to the group multiplication table, then along any given row or column, a particular element appears only once. The difference with the group multiplication table is that there, 
 along any row or column, every element is present, but in this table, not all of the elements of the set $\{ e_{ij} \}_{i,j=1}^n$ appear, because of the relation $e_{ij} e_{kl}= \delta_{jk} e_{il}$. Using the ``nice property", 
 it can be shown that the product defined in Eq.~(\ref{gubleT}) in preserved under the CJKS isomorphism. The similarity between the product defined in Eq.~(\ref{gubleT}) and the convolution defined in Eq.~(\ref{conv}) is explicit. In Eq.~(\ref{conv}), the sum is over those pairs $(x,y)$ for which the group multiplication $x \cdot y=g$, and in Eq.~(\ref{gubleT}), the sum is over those matrix units whose matrix multiplication gives $e_{ij}$. In particular, the summation is over the pairs $\{ (e_{i1},e_{1j}),(e_{i2},e_{2j}),\ldots,(e_{in},e_{nj}) \}$.    In this way, we identify the product as convolution on the space  $\mathcal{L}(\mathcal{B}(H_1),\mathcal{B}(H_2))$. 
 %
 Moreover, we have the following result.\\
 
 %
 %
 

 \noindent \textbf{Theorem.} \emph{The convolution algebra, $\mathcal{L}(\mathcal{B}(H_1),\mathcal{B}(H_2))$, is isomorphic to the algebra, $\mathcal{B}(H_1) \otimes \mathcal{B}(H_2)$, where the isomorphism is the CJKS isomorphism.}\\
 
 \noindent The proof directly follows from the discussion presented above.\\

Before going over to the next lemma, we need to discuss about matrix norms~\cite{horn_johnson_1985}. The matrix norm, $\norm{\cdot}$, on the set of $n \times n $ complex matrices is a real-valued function $\norm{\cdot}:\mathcal{M}_{n \times n}\rightarrow \mathbb{R} $ with the following constraints: 
\\

\noindent i) $\norm{A} \geq 0$. \\
ii) $\norm{A}=0$ if and only if $A=0$.\\
iii) $\norm{A+B} \leq \norm{A}+\norm{B}$. \\
iv) For any complex number $\alpha$ , $\norm{\alpha A}= \modu{\alpha} \norm{A}$. \\
v) $\norm{AB} \leq \norm{A} \norm{B}$.
\\

\noindent The conditions (i) to (iv) are just the defining conditions for a norm on a vector space. Condition (v) makes the matrix norm different from the vector space norm. Various matrix norms can be found in literature. In some cases, the $n \times n$ matrices are treated as operators on an $n$-dimensional vector space and the norm is evaluated according to the norm defined on that vector space. Such a norm is known as operator norm of matrices. Other norms include the $l_p$ norms with $p=1,2$ (for other values of $p$, the sub-multiplicative property does not hold), Schatten-\(p\) norms, etc. Here we consider only the $l_1$ and $l_2$ norms.  For a given matrix $A=(a)_{ij}$, its $l_p$ norm, $\norm{A}_{p}$, is defined as
\begin{equation}
    \norm{A}_{p}=\left(\sum_{j} \sum_{i} \modu{a_{ij}}^p \right)^{\frac{1}{p}}
    \label{pp}
\end{equation}
When $p=2$, the $l_p$ norm is known as the Frobenius norm.

In the following lemma, we associate the norm of the matrix representation of $\phi$ as the a norm of the map $\phi$. Note that the matrix representation of a map is basis-dependent, but the following lemma holds for every choice of that basis. Matrix representations of $\phi$ with respect to different bases are unitarily connected, and hence the matrix norms which are unitarily invariant give the same value for different choice of bases. The $l_2$ norm is an example of such a norm. So choosing a unitarily invariant norm for the matrix representation of $\phi$ helps us to associate a norm with $\phi$ in a basis-independent way.\\

\noindent\textbf{Lemma.} \emph{The $l_2$ norm defined on $\mathcal{L}(\mathcal{B}(H_1),\mathcal{B}(H_2))$ is sub-multiplicative with respect to convolution, i.e., $\norm{\phi_1 * \phi_2}_2 \leq \norm{\phi_1}_2 \norm{\phi_2}_2$.} 
\\

\noindent \texttt{Proof.} Let us first find the matrix element for the map $\phi_1 * \phi_2$, where we denote the matrix element of \(\phi_1\) as \((\phi_1)_{ijkl}\), and similarly for \(\phi_2\). We have
 \begin{eqnarray}
 \nonumber
      \phi_1 * \phi_2 (e_{ij}) &=& \sum_{k}\phi_1(e_{ik}) \phi_2(e_{kj}) \\
     \nonumber
     &=&\sum_{kpqrs} (\phi_1)_{pqik} e_{pq} (\phi_2)_{rskj} e_{rs}\\
     \nonumber
     &=&\sum_{kpqrs} (\phi_1)_{pqik}  (\phi_2)_{rskj} \delta_{qr} e_{ps} \\
     \nonumber
     &=&\sum_{ps}\left(\sum_{kq} (\phi_1)_{pqik} (\phi_2)_{qskj}\right) e_{ps}\\
     &=&\sum_{ps} (\phi_{12})_{psij} e_{ps}, \label{mm}
 \end{eqnarray}
 where we have denoted the matrix element of $\phi_1 * \phi_2$ as \((\phi_{12})_{psij}\), and is \(=\sum_{kq} (\phi_1)_{pqik} (\phi_2)_{qskj}\).
 Then, using the Cauchy–Bunyakovsky-Schwarz inequality, we have 
     \begin{eqnarray}
 \nonumber
    \norm{\phi_1 * \phi_2}_2 &&=\left(\sum_{ijps} \lvert (\phi_{12})_{psij} \rvert^2 \right)^\frac{1}{2} \\
     \nonumber
     &&=\left( \sum_{ijps}\lvert \sum_{kq} (\phi_1)_{pqik} (\phi_2)_{qskj} \rvert^2 \right )^{\frac{1}{2}} \\
     \nonumber
     && \leq \left( \sum_{ijps} \sum_{kq} \modu{ (\phi_1)_{pqik}}^2 \sum_{mn} \modu{(\phi_2)_{nsmj}}^2 \right )^{\frac{1}{2}} \\
     \nonumber
     && = \left( \sum_{ipkq} \modu{ (\phi_1)_{pqik}}^2 \sum_{jsmn} \modu{(\phi_2)_{nsmj}}^2 \right )^{\frac{1}{2}} \\
     \nonumber
     && = \left( \sum_{ipkq} \modu{ (\phi_1)_{pqik}}^2 \right)^{\frac{1}{2}} \left(\sum_{jsmn} \modu{(\phi_2)_{nsmj}}^2 \right )^{\frac{1}{2}} \\
     &&= \norm{\phi_1}_2 \norm{\phi_2}_2.
     \end{eqnarray}
This completes the proof. \hfill \(\blacksquare\)
\\

\noindent \textbf{Lemma.} \emph{The $l_1$ norm defined on $\mathcal{L}(\mathcal{B}(H_1),\mathcal{B}(H_2))$ is sub-multiplicative with respect to convolution, i.e., $\norm{\phi_1 * \phi_2}_1 \leq \norm{\phi_1}_1 \norm{\phi_2}_1$.}\\
 
 \noindent \texttt{Proof.} 
 %
 From Eq.~(\ref{mm}), we see that the matrix element of $\phi_1 * \phi_2$ is \((\phi_{12})_{psij}\), and is \(=\sum_{kq} (\phi_1)_{pqik} (\phi_2)_{qskj}\).
 Then, we have 
     \begin{eqnarray}
 \nonumber
\norm{\phi_1 * \phi_2}_1     &&=\sum_{ijps} \lvert (\phi_{12})_{psij} \rvert \\
     \nonumber
     &&=\sum_{ijps}\lvert \sum_{kq} (\phi_1)_{pqik} (\phi_2)_{qskj} \rvert \\
     \nonumber
     && \leq \sum_{ijps} \sum_{kq} \lvert (\phi_1)_{pqik} \rvert \lvert (\phi_2)_{qskj} \rvert \\
     \nonumber
     &&= \sum_{ijps} \sum_{kq} \sum_{m=k,n=q} \lvert (\phi_1)_{pqik} \rvert \lvert (\phi_2)_{nsmj} \rvert \\
     \nonumber
     &&\leq \sum_{ijps} \sum_{kq} \sum_{m=k,n=q} \lvert (\phi_1)_{pqik} \rvert \lvert (\phi_2)_{nsmj} \rvert \\
\nonumber     &&+ \sum_{ijps} \sum_{kq} \sum_{m \neq k,n \neq q} \lvert (\phi_1)_{pqik} \rvert \lvert (\phi_2)_{nsmj} \rvert \\
     \nonumber
     &&=\sum_{ikpq}  \lvert (\phi_1)_{pqik} \rvert \sum_{mjns} \lvert (\phi_2)_{nsmj} \rvert \\
     &&=\norm{\phi_1}_1 \norm{\phi_2}_1
 \end{eqnarray}
 This completes the proof. \hfill \(\blacksquare\)
 \\
 
 \noindent Note that the value of $\norm{\phi}_1$ depends on the choice of the basis, but for every choice of basis, the sub-multiplicative property holds.
 \\
 
\noindent \textbf{Remark.} A ``Banach algebra'' is a Banach space in which the norm is sub-multiplicative. As the convolution algebra \(\mathcal{L}(\mathcal{B}(H_1),\mathcal{B}(H_2))\) is finite-dimensional, and as the above-defined norms are sub-multiplicative, the convolution algebra is a Banach algebra  with respect to the $l_1$ and $l_2$ norms.\\


\noindent 
\textbf{Matrix representation of the convolution product.} Let the matrix forms of the two maps, $\phi$ and $\phi^\prime$, be given. Here we will discuss how to find the matrix form of the convolution these two maps.

From Eq.~(\ref{mm}), we see that $\phi_1 * \phi_2 (e_{ij})= \sum_{ps}\left(\sum_{kq} (\phi_1)_{pqik} (\phi_2)_{qskj}\right) e_{ps}$. Let $\tau$ denotes the transformation that connects Sudarshan's B-form with A-form in the situation under consideration \cite{PhysRev.121.920,2017OSID...2440016M}. The matrix form $(\phi_1)_{pqik}$  of $\phi_{1}$ is known as its A-form, and the B-form of $\phi_1$ is given by $\tau((\phi_1)_{pqik})=(\phi_1)_{piqk}$.    Eq.~(\ref{mm}) 
can be rewritten using the B-form as $\phi_1 * \phi_2 (e_{ij})= \sum_{ps}\left(\sum_{kq} {(\tau\phi_1)}_{piqk} {(\tau \phi_2)}_{qksj}\right) e_{ps}=\sum_{ps}\left( \tau{\phi_1} \tau{\phi_2}\right)_{pisj} e_{ps}=\sum_{ps}\left(\tau( \tau{\phi_1} \tau{\phi_2})\right)_{psij} e_{ps}$ . So if $\Phi_1$ and $\Phi_2$ are matrices for $\phi_1$ and $\phi_2$, then their convolution $\Phi_1 * \Phi_2$ will be of the  form,
\begin{equation}
    \Phi_1 * \Phi_2=\tau[(\tau \Phi_1) \cdot (\tau \Phi_2)],
\end{equation}
where the ``$\cdot$'' on the right-hand-side represents usual matrix product.
\\

\noindent
\textbf{Lemma.} \emph{Matrix norms which are invariant under the transformation $\tau$ is sub-multiplicative with respect to the convolution product.}\\

\noindent \texttt{Proof.} Let $\norm{\cdot}_{0}$ denote a matrix norm which is invariant under the transformation $\tau$. Then, 
\begin{eqnarray}
    \nonumber
 \norm{\Phi_1 * \Phi_2}_{0}&&=\norm{\tau[(\tau \Phi_1) \cdot (\tau \Phi_2)]}_{0} \\
    \nonumber
    &&=\norm{(\tau \Phi_1) \cdot (\tau \Phi_2)}_{0} \\
    \nonumber
    && \leq \norm{(\tau \Phi_1)}_{0} \norm{(\tau \Phi_2)}_{0} \\
    && \leq \norm{\Phi_1}_{0} \norm{\Phi_2}_{0}
\end{eqnarray} %
This completes the proof. \hfill \(\blacksquare\)
\\

\noindent Clearly the $l_1$ and $l_2$ norms are invariant under the transformation $\tau$, and hence they are sub-multiplicative with respect to the convolution.
\\

\noindent \textbf{Lemma.}
\emph{The identity of the convolution algebra, \(\mathcal{L}(\mathcal{B}(H_1),\mathcal{B}(H_2))\), is the completely depolarizing channel on $\mathcal{B}(H_1)$.}\\

\noindent \texttt{Proof.} Let $\rho$ be a arbitrary density matrix belonging to $\mathcal{B}(H_1)$. We can write $\rho$ as linear combinations of the unit matrices: $\rho = \sum_{ij} \rho_{ij} e_{ij}$. So,
    $\phi_e (\rho)= \sum_{ij} \rho_{ij} \phi_e (e_{ij}) 
    =\sum_{ij} \rho_{ij} \delta_{ij} \mathbb{I} 
    = \mbox{tr}(\rho) \mathbb{I}$. \hfill \(\blacksquare\)
    \\

\noindent\textbf{Spectrum of $\phi$.}
The spectrum of an element $\phi$ $\in$ $\mathcal{L}(\mathcal{B}(H_1),\mathcal{B}(H_2))$ can be defined as 
\begin{center}
$sp(\phi)=$  \{$\lambda \in \mathbb{C}\vert (\phi -\lambda \phi_e)$ is not invertible\}.
\end{center}
Now, due to the CJKS 
algebra isomorphism,  the non-invertibility of
$\phi - \lambda \phi_e$ is equivalent to the non-invertibility of $ \rho_{\phi}- \lambda \mathbb{I}$,  which implies that the spectrum of $\phi$ is identical to the set of eigenvalues of $\rho_{\phi}$. Here, 
$\dim{\mathcal{L}(\mathcal{B}(H_1),\mathcal{B}(H_2))}$ and $\dim{\mathcal{B}(H_1)\otimes \mathcal{B}(H_2)}$ are both finite, as \(H_1\) and \(H_2\) are assumed finite-dimensional.

 
 
 \section{Some properties of the convolution product}
 
 A quantum operation is a completely positive trace preserving (CPTP) map, and for such a map, 
 $\phi : \mathcal{B}(H_1)\rightarrow \mathcal{B}(H_2)$, its action on a density matrix $\rho$ can be 
 expressed as
 \begin{center}
     $\phi(\rho)=\sum_{{i}=1}^{n} K_i \rho  {K_i}^\dagger $,
 \end{center}
where $K_i : H_1 \rightarrow H_2$ are linear operators satisfying $\sum_{{i}=1}^{n} K_i^\dagger K_i=\mathbb{I}$. The $K_i$'s are called Kraus operators, and the above representation of $\phi$ is known as the operator-sum representation. 
This representation of a CPTP map is not unique. The sets $\{K_1, K_2,....K_n \}$ and $\{L_1, L_2,....L_m \}$  of Kraus operators represent the same CPTP map if and only if there exists a unitary operator, \(U\), such that $K_i= \sum_j U_{ij} L_j$, where the smaller set of Kraus operators is padded with zeros to make both the sets of same length. The minimum number of Kraus operators needed to represent a CPTP map is called the Kraus rank of that map. It can easily be shown that this minimum number is equal to the rank of the CJKS matrix of the CPTP map. Given a CPTP map, we can always find a representation of it with the minimum number of Kraus operators for that map, where the Kraus operators are orthogonal in the sense of the Hilbert-Schmidt inner product. If $\{ K_i \}_{i=1}^n$ is the set of Kraus operators for such a representation, then
    $\mbox{tr}(K_i^\dagger K_j) = \mbox{tr}(K_i^\dagger K_i) \delta_{ij}$.\\

\noindent\textbf{Convolution of CPTP maps.} In this paragraph, we will use the bra-ket notation, for convenience. In this notation, the matrix units $e_{ij}$ looks like $\ket{i} \bra{j}$. Let $\phi_1$ and $\phi_2$ be two CPTP maps whose Kraus operator sets with minimum number are $\{ M_p \}$ and $\{ N_q \}$ respectively.
Now the action of the convolution of $\phi_1$ and $\phi_2$ on the matrix unit, $e_{ij}$, is given by
\begin{eqnarray}
\nonumber
    \phi_1 * \phi_2 (e_{ij}) &=& \sum_k \sum_p \sum_q M_p \ket{i} \bra{k} M_p^\dagger N_q \ket{k} \bra{j} N_q^\dagger  \\
    \nonumber
   & =&\sum_p \sum_q \mbox{tr} (M_p^\dagger N_q)  M_p \ket{i} \bra{j} N_q^\dagger \\ 
   \nonumber
   & =&\sum_p \sum_q \mbox{tr} (M_p^\dagger N_q)  M_p e_{ij} N_q^\dagger.
\end{eqnarray}
 For a generic element $x \in \mathcal{B}(H_1)$ we have
 \begin{eqnarray}
 \nonumber
     \phi_1 * \phi_2 (x)=\sum_p \sum_q \mbox{tr} (M_p^\dagger N_q)  M_p x N_q^\dagger.  \end{eqnarray}
 It is  clear from the above equation that the convolution of two CP maps is, in general, not a CP map. However, if we chose $\phi_1 =\phi_2=\phi$, then 
 \begin{eqnarray}
 \nonumber
    \phi * \phi (x)=\sum_p \sum_q \mbox{tr} (M_p^\dagger M_q) \delta_{pq}  M_p x M_q^\dagger \\
     =\sum_p  \mbox{tr} (M_p^\dagger M_p)  M_p x M_p^\dagger,
 \end{eqnarray}
so that 
the convolution of a CP map with itself is a CP map and is hermiticity preserving but not trace preserving.\\

\noindent  \textbf{Unitary channels.}   The convolution of a unitary channel $\phi_U$ with itself is
     $\phi_U * \phi_U (x)= \mbox{tr} (U^\dagger U) U x U^\dagger
     =n U x U^\dagger$, 
     for all $x \in \mathcal{B}(H_1)$, where \(n = \dim H_1\), implying that $\phi_U * \phi_U=n \phi_U$.
 Conversely, if a channel $\phi$ satisfies $\phi * \phi= n \phi$, where $n$ is the dimension of the Hilbert space $H_1$, then the channel $\phi$ is a unitary channel.  
 This is the content of the following lemma.\\
 
\noindent  \textbf{Lemma.} \emph{A quantum channel $\phi : \mathcal{B}(H_1) \rightarrow \mathcal{B}(H_2)$ is a unitary channel if and only if  $\phi * \phi= n \phi$, where $n$ is the dimension of the Hilbert space $H_1$.}\\
 
\noindent \texttt{Proof.} Only one direction of the proof remains to be completed. Let $\rho_\phi$ be the CJKS matrix corresponding to the channel $\phi$. A normalized CJKS matrix  can be called the CJKS state, given by
$\rho_\phi^{CJKS} = \rho_\phi/n$. If we now assume that $\phi * \phi= n \phi$, then under the CJKS isomorphism, this implies that
     $(\rho_\phi)^2 = n \rho_\phi  
     \implies (\rho_\phi^{CJKS})^2 = \rho_\phi^{CJKS} $.
 So, $\rho_\phi^{CJKS}$ is a pure state, which implies that $\phi$ is a unitary channel. \hfill \(\blacksquare\)
 \\

\noindent \textbf{Lemma.} \emph{The convolution of transposition operation on $\mathcal{B}(H_1)$ with itself is the completely depolarizing channel.}\\

\noindent \textbf{Remark.} We have already seen that the completely depolarizing channel is the identity element of the convolution algebra. We therefore have that the transposition is the inverse of itself with respect to the convolution. \\

\noindent \texttt{Proof.} Let $T$ denote the transposition on $B(H_1)$. Its action on the matrix unit $e_{ij}$ is $T(e_{ij})=e_{ji}$. Now
$T * T (e_{ij})=\sum_k T(e_{ik}) T(e_{kj})
    =\sum_k e_{ki} e_{jk}
     =\sum_k e_{kk} \delta_{ij}= \delta_{ij} \mathbb{I}
     \implies T * T = \phi_e$. \hfill \(\blacksquare\)\\
     
 
 \noindent \textbf{Convolution of Schur multiplier maps.} Given $A,X \in \mathcal{B}(H) $, the Schur multiplier map $S_A$ associated with \(A\) is defined as $S_A(X)=A \circ X$ where $\circ$ denotes the Schur product (Hadamard product, i.e., entry-wise matrix product). The convolution of two such maps, $S_A$ and $S_B$, is given by
\begin{eqnarray}
\nonumber
    S_A * S_B(X) &=&\sum_{ij}X_{ij} \sum_k S_A (e_{ik}) S_B(e_{kj})\\
   \nonumber
    &=&\sum_{ij}X_{ij} \sum_k A\circ (e_{ik}) B \circ (e_{kj})\\
    \nonumber
    &=&\sum_{ij}X_{ij} \sum_k A_{ik} e_{ik} B_{kj}e_{kj}\\
    \nonumber
    &=&\sum_{ij}X_{ij} \sum_k A_{ik} B_{kj}e_{ij}\\
    \nonumber
    &=&\sum_{ij}X_{ij}  (AB)_{ij}e_{ij}\\
    \nonumber
    &=&(AB)\circ X \\
    &=&S_{AB}(X).
\end{eqnarray}
As $X$ is arbitrary, we have proven the following lemma.\\

\noindent\textbf{Lemma.} \emph{Given \(A,B \in \mathcal{B}(H)\), the convolution of the Schur multiplier maps, \(S_A, S_B\), is equal to the Schur multipier map associated with \(AB\):} \begin{equation}
S_A * S_B=S_{AB}.
\end{equation}
\\

\noindent \textbf{Convolution of complementary channels~\cite{doi:10.1137/S0040585X97982244}.}
Consider the Hilbert spaces $H_1, H_2, H_3$. The channel  $\phi:\mathcal{B}(H_1) \rightarrow \mathcal{B}(H_2)$ and $\Tilde{\phi} : \mathcal{B}(H_1) \rightarrow \mathcal{B}(H_3)$ are said to be complementary if there exists a linear isometry $V:H_1 \rightarrow H_2 \otimes H_3$, such that  $\phi(x)=\mbox{tr}_3 [VxV^\dagger]$ and  $\tilde{\phi}(x)=\mbox{tr}_2 [VxV^\dagger]$. Here, $V^\dagger$ denotes the adjoint of $V$. Let the channel $\phi: \mathcal{B}(H_1) \rightarrow \mathcal{B}(H_2)$ have a Kraus decomposition, $\phi(x)=\sum_{i=1}^d M_i x M_i^{\dagger}$. Then the complementary channel $\tilde{\phi}: \mathcal{B}(H_1) \rightarrow \mathcal{B}(\mathbb{C}^d)$ has the form $\tilde{\phi}(x)=\sum_{p,q=1}^d \mbox{tr} (M_q^{\dagger}M_p x)e_{pq}$.\\

\noindent \textbf{Lemma.} If $\Tilde{\phi}$ is complementary to \(\phi\), then the latter is unital 
if and only if $\tilde{\phi}*\tilde{\phi}=\tilde{\phi}$.\\

\noindent \texttt{Proof.} We begin with the if part. We have 
\begin{eqnarray} 
\nonumber
    \tilde{\phi}*\tilde{\phi} (e_{ij})&=&\sum_k \tilde{\phi}(e_{ik}) \tilde{\phi}(e_{kj}) \\
   \nonumber
    &=&\sum_{p,q,r,s=1}^d \sum_k \mbox{tr} \left(M_s^{\dagger}M_r e_{kj}\right) \mbox{tr} (M_q^{\dagger}M_p e_{ik})e_{pq} e_{rs} \\
    \nonumber
    &=& \sum_{p,q,r,s=1}^d \sum_k  \bra{j}M_s^{\dagger}M_r \ket{k} \bra{k} M_q^{\dagger}M_p \ket{i} \delta_{qr} e_{ps} \\
    \nonumber
    &=&\sum_{p,s=1}^d  \bra{j}M_s^{\dagger} \left(\sum_{q=1}^d M_q  M_q^{\dagger}\right)M_p \ket{i} e_{ps} \\
    \nonumber
    &=&\sum_{p,s=1}^d  \mbox{tr}(M_s^{\dagger}M_p e_{ij}) e_{ps}\\
   & =&\tilde{\phi} (e_{ij}).
\end{eqnarray}
The only if part can easily be proved in a similar way. \hfill \(\blacksquare\)\\

\section{Maps and map operations in  CJKS isomorphism language}


Any $\rho$ $\in $ $\mathcal{B}(H_1)\otimes \mathcal{B}(H_2) $ can be written as
    $\rho=\sum_{{i,j,k,l}=1}^{n} $ $a_{ijkl} e_{ij} \otimes e_{kl}$
    $=\sum_{{i,j}=1}^{n}$ $e_{ij}$ $\otimes$ $(\sum_{{k,l}=1}^{n}$ $a_{ijkl} e_{kl})$.
So a map $\phi \in \mathcal{L}(\mathcal{B}(H_1),\mathcal{B}(H_2)) $ can be set up for every linear operator $\rho \in \mathcal{B}(H_1)\otimes \mathcal{B}(H_2)$ by defining
\begin{equation}
    \phi(e_{ij})= \sum_{{k,l}=1}^{n} a_{ijkl} e_{kl},  
\end{equation}
extended by linearity. This is exactly the CJKS isomorphism. \\

\noindent \textbf{Partial transpose.} Let us now  define the partial transpose operation in $\mathcal{L}(\mathcal{B}(H_1),\mathcal{B}(H_2))$ via the CJKS isomorphism. We have
\begin{eqnarray}
    \rho^{T_{A}}&=&\sum_{{i,j,k,l}=1}^{n}  a_{jikl} e_{ij} \otimes e_{kl} \nonumber \\
    &=&\sum_{{i,j}=1}^{n} e_{ij} \otimes \left(\sum_{{k,l}=1}^{n} a_{jikl} e_{kl}\right).
\end{eqnarray}   
We may define the 
``partial transpose" of $\phi$ 
 by
    \begin{equation}
     \phi^{T_{A}}(e_{ij})=\sum_{{k,l}=1}^{n} a_{jikl} e_{kl}, 
    \end{equation}
extended by linearity.\\

\noindent \textbf{Trace.}
We can define the ``trace'' of an element $\phi \in \mathcal{L}(\mathcal{B}(H_1),\mathcal{B}(H_2)) $ as the sum of its spectrum, i.e., $\mbox{Tr}(\phi)=\sum_{i} \lambda_i$, with $\{ \lambda_i \}$ being the spectrum of $\phi$. Note that for this definition of trace, the relation, \begin{equation}
\mbox{Tr}(\phi_1 * \phi_2)=\mbox{Tr}(\phi_2 * \phi_1)
\end{equation}
holds for finite-dimensional 
$\mathcal{L}(\mathcal{B}(H_1),\mathcal{B}(H_2))$.  
Note that we are using ``Tr" and not ``tr'' to denote the trace.

Let $H_1=H_2=H$. The vector space $\mathcal{L}(\mathcal{B}(H_1),\mathcal{B}(H_2))$ can then be viewed as a vector space of operators acting on $\mathcal{B}(H)$. Now $\mathcal{B}(H)$ can be turned into an inner product space by the Hilbert-Schmidt inner product, $\langle A,B \rangle=\mbox{tr}({A}^\dagger B)$ for \(A, B \in \mathcal{B}(H)\). Let $\{e_{ij}\}$ be a basis for $\mathcal{B}(H)$. There is another way of defining the trace for elements of  $\mathcal{L}(\mathcal{B}(H_1),\mathcal{B}(H_2))$, viz. by treating $\phi$ as an operator acting on the $n^2$ dimensional vector space in the following way:
\begin{equation}
    \mbox{tr}(\phi)=\sum_{{i,j}=1}^{n} \langle e_{ij} , \phi(e_{ij}) \rangle.
\end{equation}
%
This 
trace can be shown to be equal to the sum of the eigenvalues of $\phi$. Therefore, 
$\mbox{tr}(\phi)$ is equal to the sum of the diagonal element of the matrix representation of $\phi$, as well as the sum of the eigenvalues of \(\phi\). 
Our definition of trace as \(\mbox{Tr}(\phi)\), may not be equal to the sum of the eigenvalues of $\phi$, because there, we are treating $\phi$ as an element of the convolution algebra $\mathcal{L}(\mathcal{B}(H_1),\mathcal{B}(H_2))$. 
The two different traces appear because the two algebras for the same set, $\mathcal{L}(\mathcal{B}(H_1),\mathcal{B}(H_2))$, happens to have different identity elements.\\


\noindent \textbf{Defining projection-like operators in $\mathcal{L}(\mathcal{B}(H_1),\mathcal{B}(H_2))$.}
Let $T$ be an operator on a vector space $V$ over the field $\mathbb{C}$, and let $\{ \lambda_i \}$ be the set of eigenvalues of  $T$. The projection operator onto the eigenspace corresponding to the eigenvalue $\lambda_i$ is given by 
%
\begin{equation}
    P_i=\prod_{j \neq i}\frac{T-\lambda_j \mathbb{I}}{\lambda_i - \lambda_j}.
\end{equation}
 Here, the product is the usual composition of two operators and $\mathbb{I}$ is the identity operator on $V$.

 To keep the similarity alive, for a given $\phi \in \mathcal{L}(\mathcal{B}(H_1),\mathcal{B}(H_2))$ and $\lambda_i$ being one of the elements of the spectrum of $\phi$, we associate with $\lambda_i$, a map $\phi_{P_i} \in \mathcal{L}(\mathcal{B}(H_1),\mathcal{B}(H_2))$
 given by 
 \begin{equation}
     \phi_{P_i}=\prod_{j \neq i}\frac{\phi-\lambda_j \phi_e} {\lambda_i - \lambda_j},
 \end{equation}
 where $\{ \lambda_i \}$ is the spectrum of $\phi$, and $\prod$ is now the convolution product.

 \section{Nonseparability of bipartite quantum operations}
 A quantum operation is a CPTP map from $\mathcal{B}(H_1)$ to $\mathcal{B}(H_2)$. The action of a CPTP map $\phi : \mathcal{B}(H_1)\rightarrow \mathcal{B}(H_2)$ on a density matrix $\rho$ can be written in terms of Kraus operators as 
     $\phi(\rho)=\sum_{{i}=1}^{n} K_i \rho  {K_i}^\dagger $
where $K_i : H_1 \rightarrow H_2$ are linear operators satisfying $\sum_{{i}=1}^{n} K_i^\dagger K_i=\mathbb{I}$.\\


\noindent \textbf{Separable bipartite quantum operations.} A bipartite quantum operation (CPTP map) $\phi : \mathcal{B}(H_1 ^A \otimes H_1 ^B ) \rightarrow  \mathcal{B}(H_2 ^A \otimes H_2 ^B ) $ is called separable if its Kraus operators are of the form
\begin{equation}
    K_i=K_i^A \otimes K_i^B 
\end{equation}
where $K_i^A : H_1^A \rightarrow H_2^A $ and $K_i^B : H_1^B \rightarrow H_2^B $ are linear operations.\\

\noindent A separable operation cannot create entangled states from unentangled ones. However, it may not be possible to implement a separable operation by local quantum operations and classical communication (LOCC)\cite{PhysRevA.59.1070}, although all LOCC are of the separable form. 


We now briefly discuss about entanglement witnesses and the Hahn Banach theorem.\\

\noindent \textbf{Hahn-Banach theorem~\cite{smith_1970, loTa-kambal}.} Let $M$ be a convex compact set in a finite-dimensional Banach space $X$. Let $\rho \notin M $ be a point in $X$. Then there exists a hyperplane that separates $\rho$ from $M$.\\

\noindent The set of separable states of a bipartite quantum system is a convex compact subset of the set of quantum states~\citep{PhysRevA.40.4277}. 
The Hahn-Banach theorem immediately guaranties the existence  of a functional  which can distinguish  any entangled state from the separable states. The linear functional can be such that  it maps the separable states to positive real numbers  and the entangled state to a negative real number. It is of the form tr$(W \cdot)$, where $W$ is a hermitian operator on the Hilbert space $H$ on which the  states under consideration act. $W$ is referred to as an ``witness operator". It has been 
shown \cite{das2017separability,RevModPhys.81.865,2009PhR...474....1G} that if the entangled state has a negative eigenvalue when partially transposed, then $W$ can be chosen as the partial transpose of the projector of an eigenvector of the partial transpose of the entangled state corresponding to a negative eigenvalue.

In the case of constructing a witness for an entangled state, we use the fact that the set of separable states is convex and closed. It can easily be shown that the  set  of  separable  operators  is  closed and convex \cite{SOHAIL2021127411}. So in a similar way, nonseparability of bipartite quantum operations can be studied and witnesses for them can be constructed.

It is shown in \cite{2001PhRvL..86..544C} that a bipartite CPTP map is separable if and only if the corresponding density operator $\rho_\phi$ under the CJKS isomorphism is separable in the bi-partition $H_1^A \otimes H_2^A : H_1^B \otimes H_2^B$. 
This 
fact and the channel-state duality can be used to study the problem of nonseparability witnesses for a bipartite quantum operation~\cite{SOHAIL2021127411}.  Nonseparability of quantum operation or ``entanglement'' of quantum operation was studied in a recent work~\cite{Chaojian} using the nonseparability of the corresponding CJKS state. They have also considered ``entanglement'' of multipartite quantum channels and constructed entanglement witnesses for circuits consisting of controlled-Z gates using the stabilizer formalism. Here we construct a witness for nonseparability of bipartite quantum operations in the space $\mathcal{L}(\mathcal{B}(H_1),\mathcal{B}(H_2))$ itself with the help of the convolution product. As the positive partial transpose criterion is applicable for detecting entanglement across any bipartition of a multipartite state, our method can be applied to witness nonseparability across any bipartition of a multipartite quantum operation.

Let $e_{ip}^{A_1}$, $e_{jq}^{B_1}$, $e_{kr}^{A_2}$, $e_{ls}^{B_2}$ be matrix units of the spaces $\mathcal{B}(H_1 ^A)$, $\mathcal{B}(H_1 ^B)$, $\mathcal{B}(H_2 ^A)$, $\mathcal{B}(H_2 ^B)$ respectively. Let $\phi : \mathcal{B}(H_1 ^A \otimes H_1 ^B ) \rightarrow  \mathcal{B}(H_2 ^A \otimes H_2 ^B ) $ be a bipartite quantum operation. The action of $\phi$ on $\mathcal{B}(H_1 ^A \otimes H_1 ^B )$ can be uniquely specified by the following equation:
\begin{equation}
    \phi(e_{ip}^{A_1} \otimes e_{jq}^{B_1})=\sum_{{k,l,r,s}=1}^{n} a_{ijklpqrs} e_{kr}^{A_2} \otimes e_{ls}^{B_2} .
\end{equation}

The partial transpose of $\rho_\phi$ in the bi-partition $H_1^A \otimes H_2^A : H_1^B \otimes H_2^B$ induces a map, say $\phi^{T_{A_1 A_2}}$ in the space $\mathcal{L}(\mathcal{B}(H_1^A \otimes H_1^B),\mathcal{B}(H_2^A \otimes H_2^B ))$ defined by the following,
\begin{equation}
    \phi^{T_{A_1A_2}} (e_{ip}^{A_1} \otimes e_{jq}^{B_1})=\sum_{{k,l,r,s}=1}^{n} a_{pjrliqks} e_{kr}^{A_2} \otimes e_{ls}^{B_2} .
\end{equation}
Now if the spectrum of $\phi^{T_{A_1A_2}}$ contains a negative real number, then $\phi^{T_{A_1A_2}}$ is guarantied to be a nonseparable bipartite quantum operation.

Let $sp(\phi^{T_{A_1A_2}})=$ $\{ \lambda_i \}$ be the spectrum of $\phi^{T_{A_1A_2}}$ and let $\lambda_i \in sp(\phi^{T_{A_1A_2}})  $ be a real number. 
The projection-like operator associated with $\lambda_i$ is 
\begin{equation}
    (\phi^{T_{A_1A_2}})_{P_{i}}= \prod_{j \neq i}\frac{\phi^{T_{{A_1A_2}}}-\lambda_j \phi_e} {\lambda_i - \lambda_j} .
\end{equation}
We can then define the witness operator for the nonseparability of the bipartite quantum operation $\phi$ as 
\begin{equation}
    W:=(\phi^{T_{A_1A_2}})_{P_{i}}^{T_{A_1A_2}}.
\end{equation}
In particular, it is  clear that  $\mbox{Tr}(W * \phi) <0$ implies that $\phi$ is nonseparable.\\

\noindent \textbf{C-NOT gate.} The action of controlled-NOT (C-NOT) gate, which we still denote by   $\phi:\mathcal{B}(H_1^A \otimes H_1^B) \rightarrow \mathcal{B}(H_2^A \otimes H_2^B) $, but where the Hilbert spaces, \(H_1^A\), \(H_1^B\), \(H_2^A\), and  \(H_2^B\) are all qubit spaces, can be expressed by the following set of equations:
\begin{center}
    $\phi(e_{00} \otimes e_{00} )=e_{00} \otimes e_{00}$, $\implies a_{00000000}=1 $, \\
    $\phi(e_{00} \otimes e_{01} )=e_{00} \otimes e_{01}$, $\implies a_{00000101}=1 $, \\
    $\phi(e_{01} \otimes e_{00} )=e_{01} \otimes e_{01}$, $\implies a_{00001011}=1 $, \\
    $\phi(e_{01} \otimes e_{01} )=e_{01} \otimes e_{00}$, $\implies a_{00001110}=1 $, \\
    $\phi(e_{00} \otimes e_{10} )=e_{00} \otimes e_{10}$, $\implies a_{01010000}=1 $, \\
    $\phi(e_{00} \otimes e_{11} )=e_{00} \otimes e_{11}$, $\implies a_{01010101}=1 $, \\
    $\phi(e_{01} \otimes e_{10} )=e_{01} \otimes e_{11}$, $\implies a_{01011011}=1 $, \\
    $\phi(e_{01} \otimes e_{11} )=e_{01} \otimes e_{10}$, $\implies a_{01011110}=1, $\\
    $\phi(e_{10} \otimes e_{00} )=e_{10} \otimes e_{10}$, $\implies a_{10110000}=1 $, \\
    $\phi(e_{10} \otimes e_{01} )=e_{10} \otimes e_{11}$, $\implies a_{10110101}=1 $, \\
    $\phi(e_{11} \otimes e_{00} )=e_{11} \otimes e_{11}$, $\implies a_{10111011}=1 $, \\
    $\phi(e_{11} \otimes e_{01} )=e_{11} \otimes e_{10}$, $\implies a_{10111110}=1 $, \\
    $\phi(e_{10} \otimes e_{10} )=e_{10} \otimes e_{00}$, $\implies a_{11100000}=1 $, \\
    $\phi(e_{10} \otimes e_{11} )=e_{10} \otimes e_{01}$, $\implies a_{11100101}=1 $, \\
    $\phi(e_{11} \otimes e_{10} )=e_{11} \otimes e_{01}$, $\implies a_{11101011}=1 $, \\
    $\phi(e_{11} \otimes e_{11} )=e_{11} \otimes e_{00}$, $\implies a_{11101110}=1, $\\
\end{center}
with the other elements of \(a\) being zero.
The corresponding map $\phi^{T_{A_1A_2}}$ can be represented by the following set of equations:
\begin{center}
    $a_{00000000}=1$, $a_{00000101}=1$, $a_{10100001}=1$, $a_{10100100}=1$, $a_{01010000}=1$,$a_{01010101}=1$,$a_{11110001}=1$, $a_{11110100}=1$, $a_{00011010}=1$, $a_{00011111}=1$, $a_{10111011}=1$, $a_{10111110}=1$, $a_{01001010}=1$, $a_{01001111}=1$, $a_{11101011}=1$, $a_{11101110}=1$,
\end{center}
and $a_{ijklpqrs}=0$ for the remaining combinations.
The action of $\phi^{T_{A_1A_2}}$ on the basis is given by 
\begin{center}
    $\phi^{T_{A_1A_2}}(e_{00} \otimes e_{00} )=e_{00} \otimes e_{00}$, 
    $\phi^{T_{A_1A_2}}(e_{00} \otimes e_{01} )=e_{00} \otimes e_{01}$, \\
    $\phi^{T_{A_1A_2}}(e_{10} \otimes e_{00} )=e_{10} \otimes e_{01}$, 
    $\phi^{T_{A_1A_2}}(e_{10} \otimes e_{01} )=e_{10} \otimes e_{00}$, \\
    $\phi^{T_{A_1A_2}}(e_{00} \otimes e_{10} )=e_{00} \otimes e_{10}$,  
    $\phi^{T_{A_1A_2}}(e_{00} \otimes e_{11} )=e_{00} \otimes e_{11}$,  \\
    $\phi^{T_{A_1A_2}}(e_{10} \otimes e_{10} )=e_{10} \otimes e_{11}$,  
    $\phi^{T_{A_1A_2}}(e_{10} \otimes e_{11} )=e_{10} \otimes e_{10}$,  \\
    $\phi^{T_{A_1A_2}}(e_{01} \otimes e_{00} )=e_{01} \otimes e_{10}$,  
    $\phi^{T_{A_1A_2}}(e_{01} \otimes e_{01} )=e_{01} \otimes e_{11}$,  \\
    $\phi^{T_{A_1A_2}}(e_{11} \otimes e_{00} )=e_{11} \otimes e_{11}$,  
    $\phi^{T_{A_1A_2}}(e_{11} \otimes e_{01} )=e_{11} \otimes e_{10}$,  \\
    $\phi^{T_{A_1A_2}}(e_{01} \otimes e_{10} )=e_{01} \otimes e_{00}$,  
    $\phi^{T_{A_1A_2}}(e_{01} \otimes e_{11} )=e_{01} \otimes e_{01}$,  \\
    $\phi^{T_{A_1A_2}}(e_{11} \otimes e_{10} )=e_{11} \otimes e_{01}$,  
    $\phi^{T_{A_1A_2}}(e_{11} \otimes e_{11} )=e_{11} \otimes e_{00}$.  \\
\end{center}
Using this data, one can calculate  $(\phi^{T_{A_1A_2}})_{P_{i}}= \prod_{j \neq i}\frac{\phi^{T_{{A_1 A_2}}}-\lambda_j \phi_e} {\lambda_i - \lambda_j} $ 
through its action on the basis $\{ e_{ij} \otimes e_{kl} \}$, and subsequently, the average
value of the operator, \(W\). \\
\\
\textbf{Swap gate:} The swap gate is another two-qubit gate,
which we again denote by   $\phi:\mathcal{B}(H_1^A \otimes H_1^B) \rightarrow \mathcal{B}(H_2^A \otimes H_2^B) $, and where again the Hilbert spaces, \(H_1^A\), \(H_1^B\), \(H_2^A\), and  \(H_2^B\) are  qubit spaces, and which can be expressed by the following equation:
%
%
\begin{eqnarray}
    \phi(e_{ip}\otimes e_{jq})=e_{jq}\otimes e_{ip}.
\end{eqnarray}
The above equation implies that the ``partial transposition'' of $\phi$ is described by the following equation:
\begin{eqnarray}
     \phi^{T_{A_1 A_2}}(e_{ip}\otimes e_{jq})=e_{qj}\otimes e_{pi}.
\end{eqnarray}
Using this equation, we can find  $(\phi^{T_{A_1A_2}})_{P_{i}}= \prod_{j \neq i}\frac{\phi^{T_{{A_1 A_2}}}-\lambda_j \phi_e} {\lambda_i - \lambda_j} $, and subsequently the witness operator for this gate.
\\
\\
\textbf{CZ gate:} The controlled-Z gate is yet another two-qubit gate,
which we yet again denote by   $\phi:\mathcal{B}(H_1^A \otimes H_1^B) \rightarrow \mathcal{B}(H_2^A \otimes H_2^B) $, and where yet again the Hilbert spaces, \(H_1^A\), \(H_1^B\), \(H_2^A\), and  \(H_2^B\) are  qubit spaces, and which can be expressed by the following set of equations:
%
%
\begin{center}
    $\phi(e_{00} \otimes e_{00} )=e_{00} \otimes e_{00}$, $\implies a_{00000000}=1 $, \\
    $\phi(e_{00} \otimes e_{01} )=e_{00} \otimes e_{01}$, $\implies a_{00000101}=1 $, \\
   $\phi(e_{00} \otimes e_{10} )=e_{00} \otimes e_{10}$, $\implies a_{01010000}=1 $, \\
    $\phi(e_{00} \otimes e_{11} )=e_{00} \otimes e_{11}$, $\implies a_{01010101}=1 $, \\
     $\phi(e_{01} \otimes e_{00} )=e_{01} \otimes e_{00}$, $\implies a_{00001010}=1 $, \\
    $\phi(e_{01} \otimes e_{01} )=-e_{01} \otimes e_{01}$, $\implies a_{00001111}=-1 $, \\
    $\phi(e_{01} \otimes e_{10} )=e_{01} \otimes e_{10}$, $\implies a_{01011010}=1 $, \\
    $\phi(e_{01} \otimes e_{11} )=-e_{01} \otimes e_{11}$, $\implies a_{01011111}=-1, $\\
    $\phi(e_{10} \otimes e_{00} )=e_{10} \otimes e_{00}$, $\implies a_{10100000}=1 $, \\
    $\phi(e_{10} \otimes e_{01} )=e_{10} \otimes e_{01}$, $\implies a_{10100101}=1 $, \\
     $\phi(e_{10} \otimes e_{10} )=-e_{10} \otimes e_{10}$, $\implies a_{11110000}=-1 $, \\
    $\phi(e_{10} \otimes e_{11} )=-e_{10} \otimes e_{11}$, $\implies a_{11110101}=-1 $, \\
    $\phi(e_{11} \otimes e_{00} )=e_{11} \otimes e_{00}$, $\implies a_{10101010}=1 $, \\
    $\phi(e_{11} \otimes e_{01} )=-e_{11} \otimes e_{01}$, $\implies a_{10101111}=-1 $, \\
    $\phi(e_{11} \otimes e_{10} )=-e_{11} \otimes e_{10}$, $\implies a_{11111010}=-1 $, \\
    $\phi(e_{11} \otimes e_{11} )=e_{11} \otimes e_{11}$, $\implies a_{11111111}=1, $\\
\end{center}
with the other elements of $a$ being zero.
The action of $\phi^{T_{A_1 A_2}}$ on the basis is given by the following set of equations:
\begin{center}
    $\phi^{T_{A_1A_2}}(e_{00} \otimes e_{00} )=e_{00} \otimes e_{00}$, 
    $\phi^{T_{A_1A_2}}(e_{00} \otimes e_{01} )=e_{00} \otimes e_{01}$, \\
    $\phi^{T_{A_1A_2}}(e_{00} \otimes e_{10} )=e_{00} \otimes e_{10}$,  
    $\phi^{T_{A_1A_2}}(e_{00} \otimes e_{11} )=e_{00} \otimes e_{11}$,  \\
    $\phi^{T_{A_1A_2}}(e_{01} \otimes e_{00} )=e_{01} \otimes e_{00}$,  
    $\phi^{T_{A_1A_2}}(e_{01} \otimes e_{01} )=e_{01} \otimes e_{01}$,  \\
    $\phi^{T_{A_1A_2}}(e_{01} \otimes e_{10} )=-e_{01} \otimes e_{10}$,  
    $\phi^{T_{A_1A_2}}(e_{01} \otimes e_{11} )=-e_{01} \otimes e_{11}$,  \\
    $\phi^{T_{A_1A_2}}(e_{10} \otimes e_{00} )=e_{10} \otimes e_{00}$, 
    $\phi^{T_{A_1A_2}}(e_{10} \otimes e_{01} )=-e_{10} \otimes e_{01}$, \\
    $\phi^{T_{A_1A_2}}(e_{10} \otimes e_{10} )=e_{10} \otimes e_{10}$,  
    $\phi^{T_{A_1A_2}}(e_{10} \otimes e_{11} )=-e_{10} \otimes e_{11}$,  \\
    $\phi^{T_{A_1A_2}}(e_{11} \otimes e_{00} )=e_{11} \otimes e_{00}$,  
    $\phi^{T_{A_1A_2}}(e_{11} \otimes e_{01} )=-e_{11} \otimes e_{01}$,  \\
    $\phi^{T_{A_1A_2}}(e_{11} \otimes e_{10} )=-e_{11} \otimes e_{10}$,  
    $\phi^{T_{A_1A_2}}(e_{11} \otimes e_{11} )=e_{11} \otimes e_{11}$.  \\
\end{center}
Using these equation, we find  $(\phi^{T_{A_1A_2}})_{P_{i}}= \prod_{j \neq i}\frac{\phi^{T_{{A_1 A_2}}}-\lambda_j \phi_e} {\lambda_i - \lambda_j} $, and subsequently we can find the witness operator for the controlled-Z gate.

\section{Conclusion}
We have defined a product in the space, $\mathcal{L}(\mathcal{B}(H_1),\mathcal{B}(H_2))$, which is preserved under the CJKS isomorphism. We have identified the defined product with the convolution product on the same space. In particular, we have shown that the convolution algebra, $\mathcal{L}(\mathcal{B}(H_1),\mathcal{B}(H_2))$, is isomorphic to the algebra $\mathcal{B}(H_1) \otimes \mathcal{B}(H_2)$, with the isomorphism being the CJKS isomorphism. This statement can be seen as  a generalization of the  theorem stating that the convolution algebra, $L(G,\mathbb{C})$, is isomorphic to the group algebra of $G$. We have also shown that with respect to the $l_1$ and $l_2$ norms, the convolution algebra $\mathcal{L}(\mathcal{B}(H_1),\mathcal{B}(H_2))$ is a Banach algebra. 
As an application of the convolution product, we have shown that it is possible to construct witness operators for nonseparability of bipartite quantum operations in the space of quantum operations itself.

\section*{Acknowledgments} 
We acknowledge useful discussions with Santanu Tantubay, Souvik Pal, and Nishant. The research of Sohail was partly supported by the INFOSYS scholarship.
We acknowledge partial support from the Department of Science and Technology, Government of India through the QuEST grant (Grant No.~DST/ICPS/QUST/Theme-3/2019/120).

\end{document}